\documentclass[sigconf]{acmart}
\newcommand{\red}[1]{\textcolor{black}{#1}}
\newcommand{\blue}[1]{\textcolor{black}{#1}}

\AtBeginDocument{%
  \providecommand\BibTeX{{%
    \normalfont B\kern-0.5em{\scshape i\kern-0.25em b}\kern-0.8em\TeX}}}

\setcopyright{rightsretained}
\acmConference[COLIEE 2021]{COLIEE 2021 workshop: Competition on Legal
   Information Extraction/Entailment }{June 21, 2021}{Online} \acmISBN{} \acmDOI{}

\begin{document}

\title{DoSSIER@COLIEE 2021: Leveraging dense retrieval and summarization-based re-ranking for case law retrieval}

\author{Sophia Althammer}
\authornote{Both authors contributed equally to this research.}
\affiliation{%
  \institution{TU Vienna}
  \city{Vienna}
  \country{Austria}
}
\email{sophia.althammer@tuwien.ac.at}

\author{Arian Askari}
\authornotemark[1]
\affiliation{
  \institution{Leiden University}
  \city{Leiden}
  \country{Netherlands}}
\email{a.askari@liacs.leidenuniv.nl}

\author{Suzan Verberne}
\affiliation{
  \institution{Leiden University}
  \city{Leiden}
  \country{Netherlands}}
\email{s.verberne@liacs.leidenuniv.nl}

\author{Allan Hanbury}
\affiliation{
  \institution{TU Vienna}
  \city{Vienna}
  \country{Austria}
}
\email{allan.hanbury@tuwien.ac.at}

\renewcommand{\shortauthors}{Althammer and Askari, et al.}

\begin{abstract}
    In this paper\blue{,} we present our approaches for the case law retrieval and the legal case entailment task in the Competition on Legal Information Extraction/Entailment (COLIEE) 2021. As first stage retrieval methods combined with neural re-ranking methods using contextualized language models like BERT \cite{bert} achieved great performance improvements for information retrieval in the web and news domain, we evaluate these methods for the legal domain. A distinct characteristic of legal case retrieval is that the query case and case description in the corpus tend to be long documents and therefore exceed the input length of BERT. We address this challenge by combining lexical and dense retrieval methods on the paragraph-level of the cases for the first stage retrieval. Here we demonstrate that the retrieval on the paragraph-level outperforms the retrieval on the document-level. Furthermore the experiments suggest that dense retrieval methods outperform lexical retrieval. For re-ranking we address the problem of long documents by summarizing the cases and fine-tuning a BERT-based re-ranker with the summaries. Overall, our best results were obtained with a combination of BM25 and dense passage retrieval using domain-specific embeddings.
\end{abstract}



\maketitle

\section{Introduction}

In case law systems, precedent cases are a primary legal resource for the decision of a new given case. Therefore it is a part of a lawyer's or paralegal's daily work to find precedent cases which support or contradict a new case \cite{turtle1995}. With the exponentially growing amount of electronically stored information in the legal domain \cite{legalrelevance}, it costs legal professionals increasingly more effort to retrieve the cases which are relevant to their case.
In order to find evidence lawyers require their search systems to find all cases which are relevant \cite{machinelearningforpriorcaseretrieval}, but at the same time, legal researchers will examine up to $50$ results in practice \cite{informationinworkplace} and therefore require a precision-oriented solution.\newline
We tackle these requirements of legal case retrieval in Task 1 of the Competition on Legal Information Extraction/Entailment (COLIEE) 2021 by first retrieving candidates from the whole corpus with the aim of a high recall and then re-ranking these candidates for precision at high ranks. In Task 2 of COLIEE 2021 it is the task to identify a paragraph of an existing case that entails the decision of a new case \cite{colieesummary} and we approach this problem with re-ranking as well.\newline
In news and web search, contextualized language models like BERT \cite{bert} brought substantial effectiveness gains to the first stage retrieval  \cite{reimers-gurevych-2019-sentence,karpukhin-etal-2020-dense,lee-etal-2019-latent,xiong2021approximate,Gao2020,hofstaetter2021efficiently} as well as to the re-ranking stage \cite{nogueira2019passage, macavaney2019cedr,hofstatter2021improving,althammer2021crossdomain}.
We aim to transfer these advantages also to the task of case law retrieval, however this task has specific challenges compared to retrieval tasks in the web and news domains: documents are written in domain specific language \cite{turtle1995}, there is a specific notion of relevance not only on document- but also on paragraph-level \cite{legalrelevance} and the documents tend to be long \cite{legalrelevance, bertpli}. For example in the case law corpus from COLIEE 2021 the documents contain on average $1274.62$ words,with minimum $1$ word and maximum $76,818$ words. This is not only the case for the candidate documents in the retrieval collection, but also for the query cases.\newline
As the input length of BERT is limited, we propose two different approaches for handling the longer documents with BERT, one for first stage retrieval and one for re-ranking. In order to achieve a high recall for the first stage retrieval, we reason that prior cases are not only relevant on a document-level to a query case, but that a prior case can be relevant to a query case only based on a single paragraph which is relevant to another paragraph of the query case \cite{bertpli,thuircoliee,westermann2020paragraph}. Therefore we propose for the first stage retrieval to split up the query case and cases in the corpus into their paragraphs and retrieve for each paragraph of the query case relevant prior cases based on the relevance of their paragraphs. For the document-level and paragraph-level retrieval, we evaluate both lexical retrieval model and of a semantic retrieval model. For the lexical retrieval model we choose BM25 \cite{bm25} and for the dense passage retrieval model we choose DPR \cite{karpukhin-etal-2020-dense}. We train the dense passage retrieval model on the legal entailing paragraph pairs as we suggest that the relevance of the paragraphs to each other is crucial for the relevance on the document-level. We denote the trained DPR model with lawDPR. When comparing the paragraph-level and document-level retrieval, we demonstrate that the paragraph-level retrieval achieves a higher recall for BM25 as well as for lawDPR and that lawDPR outperforms BM25 in terms of retrieval recall. We also show that the results of the lexical and semantic retrieval of BM25 and lawDPR complement each other and that the score combination of both retrieval models leads to superior results.\newline
For re-ranking the retrieved results we experiment with the approach of summarizing the query cases and the cases in the corpus as \citet{transummary,Rossi2019} have shown that query document summarization is valuable for case law retrieval. For this, we fine-tune Longformer Encoder-Decoder (LED) as a state-of-the-art  abstractive summarization model \cite{beltagy2020longformer} and use this for summarizing the cases. For each query and document case, we summarize the text and interpret the re-ranking as a binary classification problem and fine-tune BERT on predicting whether the summarized query case is relevant to a summarized case in the corpus or not. For Task 1 we submit 3 runs: one based on the ranking with BM25, one with the combination of BM25 and lawDPR and one with the first stage retrieval of BM25 and lawDPR and with re-ranking with BERT.\newline
For the legal entailment task (Task 2) we evaluate BM25 \cite{bm25} and the DPR model \cite{karpukhin-etal-2020-dense} trained on the entailing paragraph pairs ('lawDPR'). Here we find that the combination of the BM25 and DPR scores also improves the overall performance for identifying the entailing paragraph. For Task 2 we submit 3 runs: one based on the ranking on BM25, one based on the ranking of lawDPR and one with the combination of BM25 and lawDPR.\newline
We make the source code available at: \newline https://github.com/sophiaalthammer/dossier\textunderscore coliee.

\section{Task description}

\subsection{Task 1: The Legal Case Retrieval Task}
The aim of legal case retrieval is to design systems to automatically identify the supporting cases of a given query case, which should be noticed for solving the query case \cite{colieesummary}. The task consists of reading a new case $Q$ and selecting supporting cases $S_1, S_2, \dotsc, S_n$ ("noticed cases") from the whole case law collection for the decision of $Q$.

\subsection{Task 2: The Legal Case Entailment Task}
In the legal case entailment task, the goal is to design a system that finds paragraphs in a relevant case that entail the decision of a given new case \cite{colieesummary}. In the Task 2 of COLIEE there is a query paragraph given as well as the paragraphs of one legal case as candidates and the task is to find the paragraphs from the legal case which entail the decision of the query paragraph.

\subsection{Training and test collection}

The training and test collections for Task 1 and Task 2 contain cases from the Federal Court of Canada case laws. For Task 1, a corpus with $4415$ legal cases is given from which relevant cases should be retrieved for each query case. The cases can contain as addition to the English version also a French version. For Task 2 there are for each query paragraph the paragraphs of one case as candidates given and one needs to identify the entailing paragraphs. The statistics of the collections are in Table \ref{table:datasetstats} including the number of queries, the average number of relevant documents and the average length of the queries and candidates.

\begin{table}[]
\small
\centering
\caption{Statistics of the training and test collection for Task1 and Task2}
\begin{tabular}{@{}lcccc@{}}
\toprule
& \multicolumn{2}{c}{Task 1} & \multicolumn{2}{c}{Task 2}\\
\cmidrule(lr){2-3} \cmidrule(lr){4-5} 
           & Train & Test & Train & Test\\
\midrule
\# of queries & 650 & 250 & 425 & 100 \\
avg \# of candidates  & 4415 & 4415 & 32.12 & 32.18 \\
avg \# relevant candidates & 5.17 & 3.6 & 1.17 & 1.17 \\
avg query length (words) & 690.56 & 1817.12 & 38.26 & 37.41 \\
avg candidate length (words) & 1274.62 & 1274.62 & 102.67 & 117.91 \\
\bottomrule
\end{tabular}
\label{table:datasetstats}
\end{table}

\section{Methods}
\label{chap:methods}

\subsection{Task 1}

\begin{figure*}
    \centering
    \includegraphics{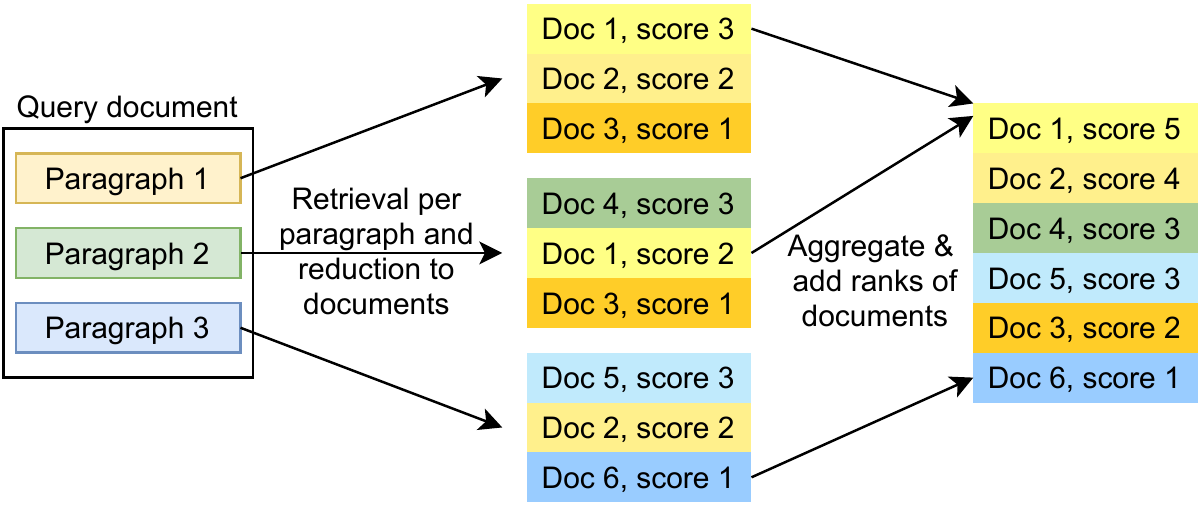}
        \caption{Aggregation of paragraph-level retrieval to an overall ranked list for the query document}
        \label{fig:aggregation}
\end{figure*}

\subsubsection{First stage retrieval}

For the lexical retrieval in the first stage retrieval we use BM25 \cite{bm25}. When querying the index with query $q$, the BM25 model assigns each document $d$ in the index a ranking score $s_{BM25}(d,q)$, the higher the score, the higher the relevance of the query and the indexed document. For semantic retrieval, we use a dense passage retrieval model \cite{karpukhin-etal-2020-dense} based on two Siamese BERT Encoders, one encodes the query passage, the other one the candidate passage. The encoder encodes the query and candidate passage into a vector and the dot-product between those two vectors denotes the relevance score $s_{lawDPR}(p,q)$ between the query $q$ and the candidate passage $p$. This dense retrieval model (lawDPR) is trained on the entailing query-paragraph pairs of Task 2, in order to align the vector representations of entailing paragraphs.\newline
As the maximum input length for the dense passage retrieval model \cite{karpukhin-etal-2020-dense} is limited, we employ this model on the full text of the case by splitting up the whole document $d$ into its paragraphs $p_1, .., p_{m_d}$ and indexing each paragraph separately. When querying the paragraph-level index, we also split up the query case $q$ into its paragraphs $q_1,.., q_{n_q}$ and retrieve for each query paragraph $q_i$ with $i \in [n_q]$ the ranked list of paragraphs $r_1,..,r_{n_q}$. The paragraph in the ranked lists $r_i$ with $i \in [n_q]$ are reduced to the paragraphs documents and then the ranked lists of documents are aggregated to one ranked list of documents for the whole query case. This aggregation will be further investigated in the next section. Equivalently we also use this approach for retrieval with BM25 on the paragraph-level and refer to this approach to \textbf{paragraph-level retrieval}.\newline
In order to analyze the impact of the paragraph-level retrieval compared to document-level retrieval on the retrieval effectiveness, we also conduct \textbf{document-level retrieval} with BM25 and lawDPR. Here all documents in the corpus are indexed based on their whole text and the query case is also encoded with all its text. As the dense passage retrieval model has a limited input length and therefore can only retrieve passages up to $512$ tokens, the document-level retrieval for lawDPR is the retrieval based on the first $512$ tokens of the query case and the candidate case.

\paragraph{Aggregation}

For the retrieval on the paragraph-level we retrieve for each paragraph of the query document $q_1,..,q_{n_q}$ a ranked list $r_1,..,r_{n_q}$ of top $N$ results. These ranked lists $r_i$ with $i \in [n_q]$ of paragraphs must be aggregated to one overall ranked list for the whole document, this process of aggregation is visualized in Figure \ref{fig:aggregation}. For the aggregation we reduce the paragraphs per ranked list $r_i$ to the paragraphs' document. As the paragraphs are reduced to their documents it is possible that one document occurs multiple times in the list. Within this reduction from the paragraph $p$ to its document $d$ the scores of the retrieval model $s(p,q_i)$ of each paragraph $p$ in the ranked list $r_i$ are replaced with an integer scores of the document $d$ $s(d,q_i)$ according to their position: The first document gets the score $N$, per position in the list the score decreases by $1$ so that the last document gets the score $1$. Then we aggregate the lists $r_i$ per query paragraph $q_i$ to one ranked list for the whole query document by adding the integer scores for each retrieved document and ranking by the new additive scores. The new additive score of a document $d$ for a query $q$ is then:

\begin{align*}
    s(d,q) = \sum_{i=1}^{n_q} \sum_{d \in r_i} s(d,q_i)
\end{align*}

We also experiment with other aggregation strategies using the similarity scores of the retrieval model or using an interleaving approach, but find that the aggregation with the additive integer scores $s(d,q)$ for a document $d$ and query $q$ lead to the overall best performance.

\paragraph{Combination}
We find that $67\%$ of the relevant cases of Top1000 which are retrieved from lawDPR on paragraph-level and $70\%$ of the relevant cases of Top1000 which are retrieved from BM25 on paragraph-level are the same documents. Therefore $23\%$ of the retrieved relevant cases from lawDPR and $30\%$ of the retrieved relevant cases from BM25 are different, hence BM25 and lawDPR retrieve different relevant documents.
For the first stage retrieval we combine for each document $d$ in the index the relevance scores of BM25 $s_{BM25}(d,q)$ and the relevance score for the whole document $d$ of lawDPR $s_{lawDPR}(d,q)$ from paragraph-level retrieval with BM25 and lawDPR.
In order to get an overall score of a document $d$ and a query $q$ we add the BM25 and the lawDPR relevance scores with a scalar weighting $\alpha, \beta \in \mathbb{R}$

\begin{align*}
    s_{BM25+lawDPR}(d,q) = \alpha \ s_{BM25}(d,q) + \beta \ s_{lawDPR}(d,q).
\end{align*}

We denote the combination with BM25+lawDPR.

\subsubsection{Re-ranking}

\paragraph{Summarizer: LED}
The current state of the art in abstractive summarization is Transformer models \cite{lewis2019bart,raffel2019exploring}. However, the input of pre-trained available models of these architectures is limited to 1024 tokens, and the majority of case law documents in our collection is longer than that. \citet{beltagy2020longformer} proposed Longformer-Encoder-Decoder (LED), which is a Transformer variant that supports much longer inputs. For this competition, we evaluate the effectiveness of LED for case law retrieval.

\paragraph{BERT}
For re-ranking we use BERT and fine-tune the pre-trained BERT model (BERT-Base, uncased) with a linear combination layer stacked atop the classifier [CLS] token on binary classification if a query summmary and a document summary are relevant to each other or not. We represent the query as sentence A and the document as sentence B in the BERT input: \begin{equation*}
    ``[CLS] \enspace query \enspace document \enspace [SEP] \enspace candidate \enspace document \enspace [SEP]"
\end{equation*}

\paragraph{Cut-off value}
As the task is to retrieve the relevant cases to a given query $q$, we consider the top-k-ranked documents $d$ in the ranked list as relevant and denote $k$ as cut-off value. We evaluate the best cut-off value $k$ depending on the F1-score of the validation set.

\subsection{Task 2}

For identifying the entailing paragraphs $p$ to a given query paragraph $q$ we use lexical and semantic ranking approaches in order to rank the given candidate paragraphs $p$. In order to predict which paragraphs of the ranking are entailing the query paragraph, we consider the top-k-ranked paragraphs, where we denote $k$ as cut-off value. We evaluate the best cut-off value $k$ depending on the F1-score on the validation set.\newline
In order to rank each paragraph given the query paragraph, we create per query paragraph an index containing the given candidate paragraphs. Then we query this index with the query paragraph and obtain the ranking for the candidate paragraphs. We do this procedure either using BM25 or lawDPR in order to compare both models for the legal entailment task.

\paragraph{BM25} For the lexical ranking we use BM25 \cite{bm25}.

\paragraph{lawDPR} For the semantic retrieval we use the same dense retrieval model \cite{karpukhin-etal-2020-dense} as in Task 1, which is trained on the entailing query-paragraph pairs of the training collection for Task 2. As Task 2 is on paragraph-level we can directly use the dense passage retrieval model without aggregating the results as in Task 1.

\paragraph{BM25+lawDPR} We also combine the ranking of BM25 and lawDPR with the same method as for Task 1.

\section{Experiment design}
\label{chap:experiments}

\subsection{Data pre-processing}

For the experiments we divide the training collections of Task 1 and Task 2 into a training and validation set. The validation sets for Task 1 and for Task 2 contain the last 100 query cases of the training sets, respectively.\newline
For the data pre-processing for Task 1, we remove the French versions of the cases when reading in the cases. We divide the cases into an introductory part, a summary, if they contain one, and their paragraphs. As we want to distinguish the cases by their text, we remove text parts of the cases which appear exactly the same in multiple cases, as they do not add any information which is particular for one case. Therefore we remove introductory parts and summaries, which appear exactly the same in more than $100$ cases from the case text as we consider them as non-informative. The introductory parts have an average length of $73.66$ words, the summaries have on average $227.05$ words and the paragraphs have a average length of $92.80$ words. We use the same data pre-processing for all submitted runs.\newline
We pre-process the data of Task 2 by reading in the paragraphs, splitting the words at whitespaces and removing tabs. Here we also use the same data pre-processing for all submitted runs.

\newpage

\subsection{Task 1}

\subsubsection{First stage retrieval}

\paragraph{BM25} We use the BM25 implementation from ElasticSearch\footnote{https://github.com/elastic/elasticsearch} with default parameter values $k=1.2$ and $b=0.75$.

\paragraph{lawDPR} We train the dense retrieval model with two BERT-based-uncased Siamese Encoders on the training set and validation set of Task 2 in the same fashion as in \cite{karpukhin-etal-2020-dense}. Here a pair of query and candidate paragraph which are relevant to each other are considered as positive sample and a pair of query and candidate paragraph which are not relevant to each other are considered as negative sample. Different to \citet{karpukhin-etal-2020-dense} we sample the negative candidate paragraphs to a query paragraph randomly from the paragraphs which are not denoted as relevant to this query paragraph. We train the model for $40$ epochs with a pairwise loss for the first $30$ epochs and a list-based loss for the last $10$ epochs, in order to include the pairwise and listwise relation of the samples in the training. We use batch size of $22$, a maximum sequence length of $256$ and a learning rate of $2*10^{-5}$.

We index the corpus on the paragraph- and on the document-level using BM25 and lawDPR and then query each index with the query cases. For the paragraph-level retrieval the query is split up into its paragraphs, for each query paragraph a ranked list of paragraphs is retrieved and aggregated to a ranked list of documents as described in section \ref{chap:methods}. For the first stage retrieval on the paragraph-level the introductory part and the summary are also treated as paragraph of the document. For the document-level retrieval the documents are indexed based on their whole text and for each query case the whole text is taken into account.

\paragraph{BM25+lawDPR} We experiment with combining the scores of BM25 and lawDPR as described in section \ref{chap:methods}. We take the Top1000 aggregated lists for the paragraph-level retrieval of BM25 and lawDPR and use the weights $ [ \alpha , \beta ] \in \{[1,1], [2,1], [3,1],$  $[4,1]\}$.

\subsubsection{Re-ranking}

\paragraph{Summarizer: LED}
For LED, we used the Huggingface transformers
library\footnote{https://huggingface.co/transformers/} and set the local attention window size to 512 tokens. To limit memory use, we use gradient checkpointing and set the input size in training to $8192$ tokens which covers more than $86\%$ of COLIEE'18 Task 1 documents completely (the longer documents are truncated at $8912$ tokens). 
We set the maximum length to generate a summary for an unseen document as $10\%$ of the length of the original text. 
We use the COLIEE'18 Task 1 data, which contains human reference summaries, for evaluation of the summarizers\footnote{As opposed to COLIEE’19 and COLIEE’20, COLIEE’18 contains expert-written summaries.}.
\par
In COLIEE'18, the summaries are provided for all queries and more than $80\%$ of document cases. We extracted summaries like \cite{tran2020encoded}, and after removing duplicates, $6,257$ unique documents are left for which a summary is available. Finally, we fine-tuned the LED model on the unique documents that contains summary. We kept the other hyperparameters (optimizer, dropout, weight decay) identical to \cite{beltagy2020longformer} and set the global attention on the first <s> token.
We use the fine-tuned LED on COLIEE'18 to generate summary for query and document cases in COLIEE'21, and provide the summaries as the representation of cases in BERT-based re-ranking step.
\paragraph{BERT-based re-ranking}
We truncate the summarized documents such that the concatenated query document (\red{truncated at} 100 words), candidate document, and the separator tokens do not exceed $512$ tokens. We re-rank the top-500 BM25+lawDPR results. We find $k=7$ as the optimal cut-off of the ranking for the selection of documents. We train each model for $100$ epochs, each with $32$ batches of $16$ training pairs, with the initial learning rate of $3*10^{-5}$, followed by a power 3 polynomial decay.

For Task 1 we submit 3 runs: one based on the paragraph-level retrieval and ranking of BM25, the second based on the combination of paragraph-level retrieval of BM25 and lawDPR and the third takes the retrieved candidates from the combination of BM25 and lawDPR and re-ranks them using the summarizer and BERT.

\subsection{Task 2}

\paragraph{BM25} We use the BM25 implementation from ElasticSearch with default parameters and create an index for each query paragraph and its candidates.

\paragraph{lawDPR} We take the trained lawDPR model from Task 1.

\paragraph{BM25+lawDPR} We combine the scores of BM25 and lawDPR as described in scetion \ref{chap:methods} and use the weights $ [ \alpha , \beta ] \in \{[1,1], [2,1], [3,1]$\\$, [4,1]\}$.
For Task 2 we submit 3 runs: one based on BM25, the second based on lawDPR and the third with the combination of BM25 and lawDPR.

\begin{table}[t]
\small
\centering
\caption{Task 1: Recall@N for the first stage retrieval of BM25 and lawDPR on the paragraph-level and the document-level retrieval on validation set}
    \begin{tabular}{@{}lccccccc@{}}
    \toprule
    Model & Retrieval level  &R@100 & R@200 &R@300  & R@500 \\
    \midrule
    BM25 & document-level & 0.5519 &0.6525	&0.6938	&0.7715	\\
    BM25 & paragraph-level & 0.5303	& 0.6765 &0.7469	&0.8164\\
    lawDPR & document-level & 0.1172	&0.1381 &0.1651	&0.2466	 \\
    lawDPR & paragraph-level & 0.4776	&0.6111	& 0.6757  &0.7735 \\
    BM25+lawDPR & paragraph-level & \textbf{0.5639}	& \textbf{0.6932}	&\textbf{0.7577} &	\textbf{0.8193}\\
    \bottomrule
    \end{tabular}
    \label{table:evalfirststagerretrieval}
\end{table}

\begin{table}[t]
\small
\centering
\caption{Task 1 validation set evaluation}
    \begin{tabular}{@{}lccc@{}}
    \toprule
    Run&Precision & Recall&F1-Score \\
    \midrule
    BM25 (cutoff at 7) &0.0948 & \textbf{0.0471} & 0.0629 \\
    BM25+lawDPR (cutoff at 7) &  \textbf{0.0987} & \textbf{0.0471}  &\textbf{0.0638}\\
    BERT re-ranking (cutoff at 7) &0.0296 & 0.0157 & 0.0205\\
    \bottomrule
    \end{tabular}
    \label{table:evaltask1}
\end{table}

\begin{table}[t]
\small
\centering
\caption{Task 1 test set evaluation}
    \begin{tabular}{@{}lccc@{}}
    \toprule
    Run&Precision & Recall&F1-Score \\
    \midrule
    BM25 (cutoff at 7) & \textbf{0.0777} & \textbf{0.1959} & \textbf{0.1113} \\
    BM25+lawDPR (cutoff at 7) &  0.0737 & 0.1788 & 0.1044\\
    BERT re-ranking (cutoff at 7) &0.0211  & 0.0546 & 0.0304 \\
    \bottomrule
    \end{tabular}
    \label{table:evaltask1_test}
\end{table}

\section{Results and analysis}

\subsection{Task 1}

\subsubsection{First stage retrieval}

We compare the first stage retrieval of BM25 and lawDPR on the paragraph- and the document-level and also evaluate the combination of the scores of BM25 and lawDPR. In the first stage retrieval it is our goal to achieve a high recall, therefore we evaluate the recall@100, recall@200, recall@300 and recall@500 using pytrec\textunderscore eval\footnote{https://github.com/cvangysel/pytrec\textunderscore eval} on the validation set. The results are shown in Table \ref{table:evalfirststagerretrieval}. As described in section \ref{chap:experiments} we combine the scores of BM25 and lawDPR and obtain a list ranked by the combination of the BM25 and lawDPR scores. When evaluating the retrieval on the validation set, we find that the combination with the weights $[3,1]$ leads to the best results, therefore we only present the results for this weighting and denote it as BM25+lawDPR.

Comparing the retrieval on paragraph- and document-level we see that the retrieval on paragraph-level outperforms the retrieval on whole document-level for BM25 as well as for lawDPR. This shows that our approach of tackling the long documents for a contextualized language model with limited input length is not only beneficial for the dense passage retrieval model, but also for the lexical retrieval with BM25.\newline
Furthermore we see that BM25 outperforms the first stage retrieval on paragraph-level of lawDPR by $4-6\%$ in terms of recall@N. However when combining the scores of BM25 and lawDPR, we see that the overall recall is the best at all evaluated cut-off values.

\subsubsection{Ranking}

\begin{figure}
    \centering
    \includegraphics[width=0.5\textwidth]{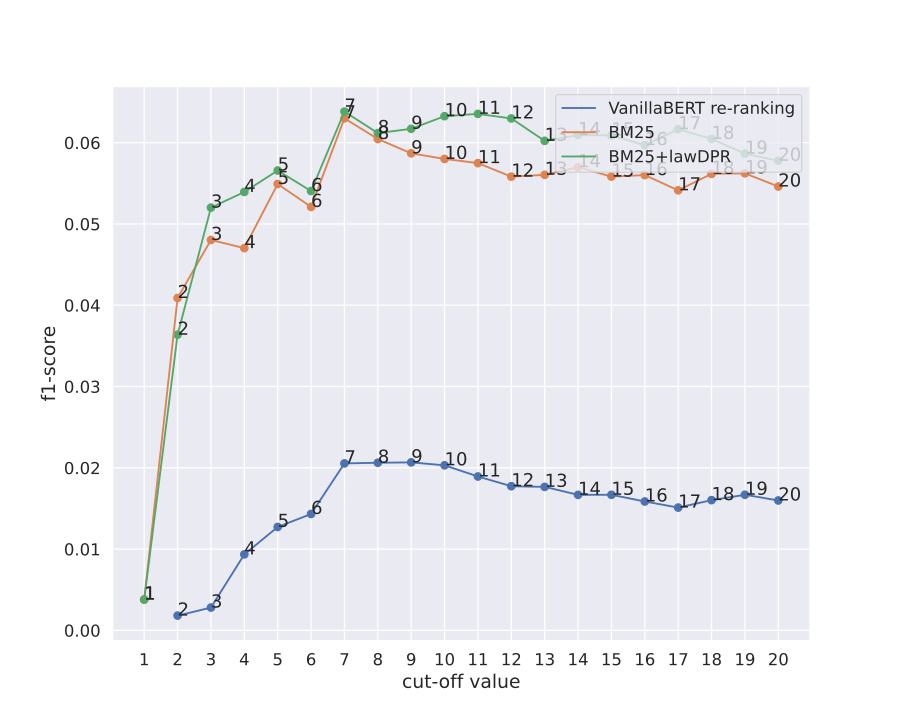}
        \caption{F1-score for Task 1 on the validation set for different cut-off values of our runs}
        \label{fig:evaltask1}
\end{figure}

We evaluate the precision, recall and F1-scores for the runs we submitted at the cut-off value of $7$ for each run as we find in out analysis in Figure \ref{fig:evaltask1} that the best cut-off value in terms of F1-score is $7$ for all three runs. The evaluation results on the validation set are in Table \ref{table:evaltask1} and on the test set in Table \ref{table:evaltask1_test}.\newline
Here we see that the overall ranking performance is improved for the validation set of BM25+lawDPR, this approach also outperforms the re-ranking based on the summaries. Contrary to that we find that on the test set BM25 achieves the best performance in terms of precision, recall and F1-score.\newline
There is a difference in the evaluation scores between our evaluation and the evaluation of the task coordinator. We use for our evaluation the off-the-shelf pytrec\textunderscore eval\footnote{https://github.com/cvangysel/pytrec\textunderscore eval} library and were in contact with the task organizer, however we could not clarify the differences between the evaluations.

\begin{figure}
    \centering
    \includegraphics[width=0.5\textwidth]{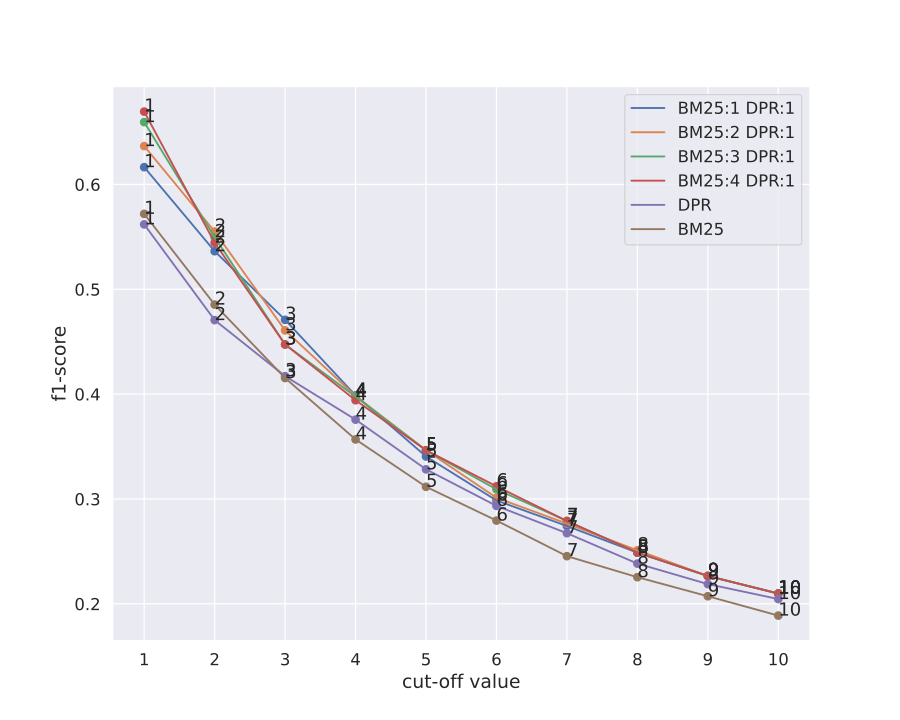}
        \caption{F1-score of task 2 on the validation set for different cut-off values of our runs, BM25:$\alpha$ DPR: $\beta$ denotes the weighting factors of the BM25+lawDPR run}
        \label{fig:evaltask2f1}
\end{figure}

\begin{table}[t]
\small
\centering
\caption{Task 2 validation set evaluation}
    \begin{tabular}{@{}lccc@{}}
    \toprule
    Run&Precision & Recall&F1-Score \\
    \midrule
    BM25 (cutoff at 1) & 0.5583 & 0.6300 &0.5719\\
    lawDPR  (cutoff at 1) &0.5283 & 0.6000 & 0.5618\\
    BM25+lawDPR (cutoff at 1) &  \textbf{0.6333} & \textbf{0.7100}  & \textbf{0.6694}\\
    \bottomrule
    \end{tabular}
    \label{table:evaltask2}
\end{table}

\subsection{Task 2}

\begin{figure}
    \centering
    \includegraphics[width=0.5\textwidth]{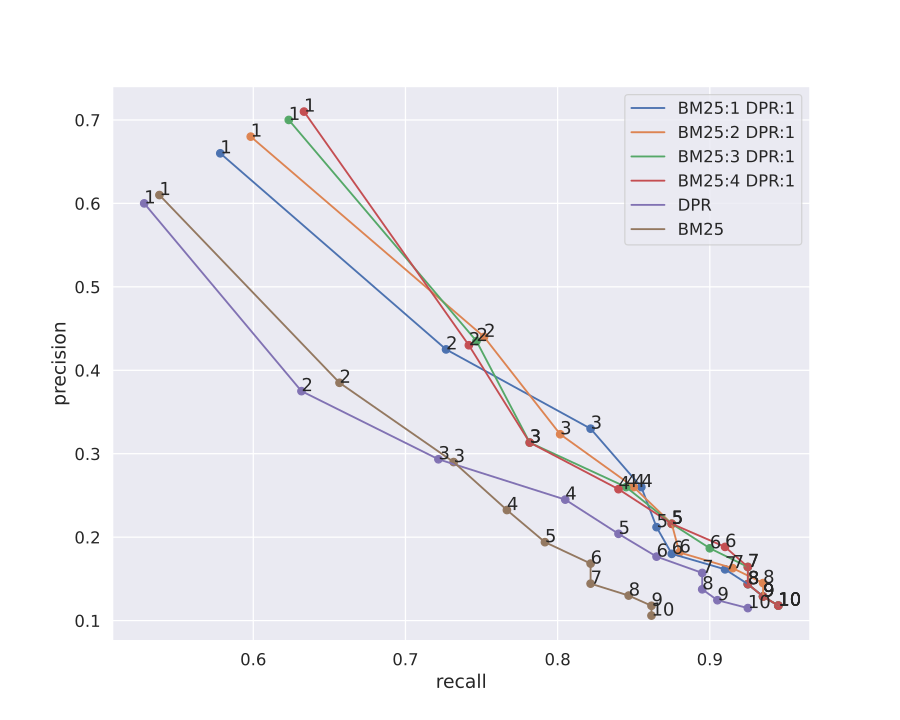}
        \caption{Precision and recall of Task 2 on the validation set for different cut-off values of our runs, BM25:$\alpha$ DPR: $\beta$ denotes the weighting factors of the BM25+lawDPR run}
        \label{fig:evaltask2precrec}
\end{figure}

For Task 2 we evaluated precision, recall and F1-score on the evaluation set for multiple cut-off values. The overall F1-score for multiple cut-off values is visualized in Figure \ref{fig:evaltask2f1}, the performance for the precision and recall of various cut-off values is visualized in Figure \ref{fig:evaltask2precrec}.
Figure \ref{fig:evaltask2precrec} shows how the precision decreases with an increasing cut-off value and therefore an increasing recall. Here we also see that BM25 and lawDPR have a similar performance, with lawDPR having a better performance with a higher cut-off value than BM25. We also clearly see the gap between BM25 and lawDPR and the combination of the scores of BM25 and lawDPR. We see a similar picture in Figure \ref{fig:evaltask2f1}, here the F1-score is continuously decreasing with a higher cut-off value, therefore we select a cut-off value of $1$ for each run. We also see that the overall performance is the best with the weight $[4,1]$ therefore we use these weights for combining BM25 and lawDPR.\newline
In Table \ref{table:evaltask2} we can see the  precision, recall and F1-score on the validation set for BM25, lawDPR and BM25+lawDPR. Here we see that BM25 has a better performance than lawDPR, but that both runs are outperformed by the combination of BM25 and lawDPR. This also suggests that combining the strengths of lexical and semantic retrieval models is beneficial for legal paragraph entailment and that BM25 and lawDPR complement each other.\newline
We also evaluate our three final runs on the test set, the evaluation results can be found in Table \ref{table:evaltask2_test}. Here we find the same relations between the evaluation results as for the validation set evaluation.

\begin{table}[t]
\small
\centering
\caption{Task 2 test set evaluation}
    \begin{tabular}{@{}lccc@{}}
    \toprule
    Run&Precision & Recall&F1-Score \\
    \midrule
    BM25 (cutoff at 1) & 0.6300 & 0.5953 & 0.6121 \\
    lawDPR  (cutoff at 1) & 0.5600 & 0.5203 &  0.5394\\
    BM25+lawDPR (cutoff at 1) &  \textbf{0.7100} & \textbf{0.6753} & \textbf{0.6922}\\
    \bottomrule
    \end{tabular}
    \label{table:evaltask2_test}
\end{table}

\section{Conclusion and future work}

Our participation at COLIEE 2021 in Task 1 and Task 2 gave the opportunity to explore information retrieval challenges in the legal case law retrieval and legal entailment. Our goal is to combine traditional lexical retrieval models with dense passage retrieval models as well as use contextualized re-ranking models for re-ranking the results. For legal case retrieval we identify the challenge of long documents for dense retrieval and neural re-ranking strategies. Therefore we present, compare and evaluate two approaches for handling the long documents:
\begin{itemize}
    \item Splitting up the documents into their paragraphs and proposing a paragraph-level retrieval for first stage retrieval and
    \item Generating summaries for the neural re-ranking with BERT.
\end{itemize}
We show that the paragraph-level retrieval in the first stage outperforms the document-level retrieval and that the combination of lexical and semantic retrieval models leads to the best results. Also for the ranking we find that the combination of lexical and semantic models improves the overall effectiveness of the ranking. Our experiments with re-ranking based on the summaries of the cases seems not to improve the overall ranking effectiveness.
Furthermore we also find that the combination of lexical and semantic retrieval methods improves the overall performance also for the passage entailment task. This leads to the conclusion that lexical and semantic retrieval methods have different strengths and complement each other.
In future we plan to investigate dense retrieval models with domain specific contextualized language models like LegalBERT \cite{chalkidis-etal-2020-legalbert} and with different paragraph aggregation approaches.

\begin{acks}
This work was supported by the EU Horizon 2020 ITN/ETN on Domain Specific Systems for Information Extraction and Retrieval (H2020-EU.1.3.1., ID: 860721).
\end{acks}

\bibliographystyle{ACM-Reference-Format}
\bibliography{sample-base}



\end{document}